\documentclass[sigconf]{acmart}

\AtBeginDocument{%
  \providecommand\BibTeX{{%
    \normalfont B\kern-0.5em{\scshape i\kern-0.25em b}\kern-0.8em\TeX}}}

\AtBeginDocument{%
 \providecommand\BibTeX{{%
    \normalfont B\kern-0.5em{\scshape i\kern-0.25em b}\kern-0.8em\TeX}}}
    
\usepackage{enumitem}
\usepackage{multirow}
\usepackage{makecell}
\usepackage{amsthm}
\usepackage{amsfonts}
\usepackage{url}

\newcommand{\ie}{\emph{i.e., }}
\newcommand{\eg}{\emph{e.g., }}
\newcommand{\etal}{\emph{et al. }}

\newcommand{\etc}{\emph{etc.}}
\newcommand{\wrt}{\emph{w.r.t. }}
\newcommand{\cf}{\emph{cf. }}

\newcommand{\wx}[1]{{\color{black}{#1}}}
\newcommand{\za}[1]{{\color{black}{#1}}}

\copyrightyear{2022}
\acmYear{2022}
\setcopyright{rightsretained}

\acmConference[KDD '22]{Proceedings of the 28th ACM SIGKDD Conference on Knowledge Discovery and Data Mining}{August 14--18, 2022}{Washington, DC, USA}
\acmBooktitle{Proceedings of the 28th ACM SIGKDD Conference on Knowledge Discovery and Data Mining (KDD '22), August 14--18, 2022, Washington, DC, USA}
\acmISBN{978-1-4503-9385-0/22/08}
\acmDOI{10.1145/3534678.3539229}

\usepackage{etoolbox}
\makeatletter
\patchcmd{\maketitle}{\@copyrightpermission}{
   \begin{minipage}{0.3\columnwidth}
     \href{https://creativecommons.org/licenses/by/4.0/}{\includegraphics[width=0.90\textwidth]{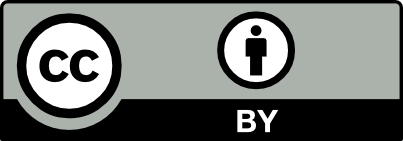}}
   \end{minipage}\hfill
   \begin{minipage}{0.7\columnwidth}
     \href{https://creativecommons.org/licenses/by/4.0/}{This work is licensed under a Creative Commons Attribution International 4.0 License.}
   \end{minipage}

   \vspace{5pt}
}{}{}

\makeatother

\begin{document}

\title{CrossCBR: Cross-view Contrastive Learning for Bundle Recommendation}

\author{Yunshan Ma}
 \affiliation{
     \institution{Sea-NExT Joint Lab \& National University of Singapore}
     \country{}
 }
 \email{yunshan.ma@u.nus.edu}
 
 \author{Yingzhi He}
 \country{}
 \affiliation{
    \institution{Sea-NExT Joint Lab \& National University of Singapore}
    \country{}
 }
 \email{heyingzhi@u.nus.edu}

 \author{An Zhang}
 \authornote{Corresponding author.}
 \affiliation{
    \institution{Sea-NExT Joint Lab \& National University of Singapore}
    \country{}
 }
 \email{an\_zhang@nus.edu.sg}

 \author{Xiang Wang}
 \affiliation{
    \institution{University of Science and Technology of China}
    \country{}
 }
 \email{xiangwang1223@gmail.com}

 \author{Tat-Seng Chua}
 \affiliation{
     \institution{Sea-NExT Joint Lab \& National University of Singapore}
     \country{}
 }
 \email{dcscts@nus.edu.sg}

\begin{abstract}
Bundle recommendation aims to recommend a bundle of related items to users, which can satisfy the users' various needs with one-stop convenience. Recent methods usually take advantage of both user-bundle and user-item interactions information to obtain informative representations for users and bundles, corresponding to bundle view and item view, respectively. However, they either use a unified view without differentiation or loosely combine the predictions of two separate views, while the crucial cooperative association between the two views' representations is overlooked. 

In this work, we propose to model the cooperative association between the two different views through cross-view contrastive learning. By encouraging the alignment of the two separately learned views, each view can distill complementary information from the other view, achieving mutual enhancement. Moreover, by enlarging the dispersion of different users/bundles, the self-discrimination of representations is enhanced. Extensive experiments on three public datasets demonstrate that our method outperforms SOTA baselines by a large margin. Meanwhile, our method requires minimal parameters of three set of embeddings (user, bundle, and item) and the computational costs are largely reduced due to more concise graph structure and graph learning module. In addition, various ablation and model studies demystify the working mechanism and justify our hypothesis. Codes and datasets are available at \url{https://github.com/mysbupt/CrossCBR}.
\end{abstract}


\maketitle
\renewcommand{\shortauthors}{Yunshan Ma et al.}

\section{Introduction} \label{sec:introduction}

Bundle recommendation aims to recommend a \za{set} of \wx{items related with the same theme to users}. 
In \za{a variety of} online applications, such as music platforms and fashion shopping \za{sites}, serving bundles instead of individual items can \za{boost the users' experience in a one-stop manner.}
More importantly, \za{platforms taking bundles as the marketing strategy can increase sales revenue and attract customers fond of bundle discounts.}
Consequently, both the users and platforms \wx{would prefer bundles} (\eg music \za{playlist} and fashion outfit) instead of single items (\eg single \za{song} and \za{piece-of-clothing}).
Therefore, developing effective bundle recommender systems \wx{is attracting a surge of interest in both academia and industry.}

\begin{figure}[!t]
    \centering
    \includegraphics[width = 0.8\linewidth]{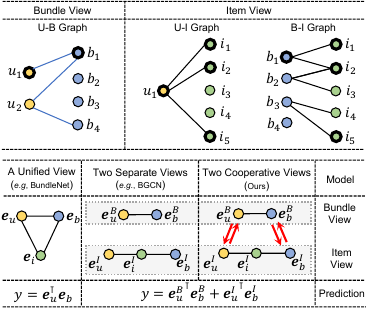}
    \vspace{-6pt}
    \caption{Top: The bundle and item views presented in the U-B, U-I and B-I graphs. Bottom: Our work models the cooperative association between views, where the superscripts $B$ and $I$ denote the bundle and item view,  and the subscripts $u$, $b$, and $i$ stand for the user, bundle, and item.}
    \label{fig:motivation}
    \vspace{-12pt}
\end{figure}

\wx{Scrutinizing prior studies on bundle recommendation \cite{rendle2010factorizing,cao2017embedding,chen2019matching,deng2020personalized,chang2020bundle}, we can systematize the sources of user preferences as two views:
(1) bundle view, which depicts user preferences through the \underline{u}ser-\underline{b}undle interactions and can be reorganized as an U-B graph;
and (2) item view, which delineates user behaviors and bundle knowledge at the granularity of items --- \ie \underline{u}ser-\underline{i}tem interactions in the form of an U-I graph and \underline{b}undle-\underline{i}tem affiliation in the form of a B-I graph, respectively.
These two views allow us to understand user interests and construct the recommender models from different perspectives. However, there exist clear discrepancies between these two views which have not been modeled in prior studies.
Consider the running example on the top part of Figure \ref{fig:motivation}, where $u_1$ is the target user who has interacted with bundle $b_{1}$ and individual items $i_1$, $i_2$, and $i_5$ before, and the historical bundle $b_{1}$ consists of items $i_1$ and $i_2$.
By taking the bundle view solely, a recommender model is highly likely to route $b_4$ to $u_1$, if the behavioral similarity between users $u_1$ and $u_2$ is well captured.
On the other hand, by taking the item view, a recommender model is prone to yield $b_2$ and $b_3$ as the interested item of $u_1$, since $b_2$ holds items (\ie $i_2$) shared with the historical bundle $b_1$ and $b_3$ contains items (\ie $i_5$) individually preferred by $u_1$.
Clearly, the bundle view emphasizes the behavioral similarity among users, while the item view highlights the content relatedness among bundles and users' item-level preference.
Hence they are complementary but different, and the cooperation of these two views is the key to accurate bundle recommendation.
}

However, the cooperative association \wx{between these two views} has been loosely modeled or even overlooked in \za{existing} works \cite{cao2017embedding,chen2019matching, deng2020personalized,chang2020bundle}.
\wx{One research line like BundleNet \cite{deng2020personalized} blindly merges the two views into a unified tripartite graph and employs graph neural networks (GNNs) to aggregate the neighboring information into representations of users and bundles.
However, such representations fail to differentiate the behavioral similarity among users and content relatedness among bundles from these two views, thus obscuring their cooperation.
Another line, such as BGCN \cite{chang2020bundle}, first performs representation learning and preference prediction upon the views individually, and then fuses these two view-specific predictions.
While this loose modeling of the two views performs better than the first line, it only considers the cooperative signal at the level of predictions, rather than directly plugging such signal into the representations optimized for recommendation.
Hence, no mutual enhancement of the two views is guaranteed to be captured.
Considering the limitations of the two lines, we believe that it is crucial to properly model the cooperative association and encourage the mutual enhancement across the bundle and item views.
}


\wx{Towards this end,} we propose a \textbf{Cross}-view \textbf{C}ontrastive Learning for \textbf{B}undle \textbf{R}ecommendation (CrossCBR) \wx{which captures the cooperative association by cross-view contrastive learning and mutually enhances the view-aware representations.
The basic idea is to treat the bundle and item views as two distinct but correlated viewpoints of user-bundle preferences, and apply contrastive learning on these viewpoints to encapsulate their agreements into representations of users and bundles.
Specifically, upon the U-B graph, we build a LightGCN \cite{he2020lightgcn} as the backbone to obtain the bundle-view representations of users and bundles;
analogously, upon the U-I graph, we employ another LightGCN to generate the item-view representations of users and items, and aggregate the representations of compositional items as the bundle representation based on the B-I graph.
We jointly employ the BPR \cite{rendle2012bpr} and contrastive loss \cite{gutmann2010noise} to optimize these representations.
Benefiting from the cross-view contrastive learning, CrossCBR outperforms the state-of-the-art (SOTA) baselines by a large margin on three datasets.}


To demystify the working mechanism behind \wx{CrossCBR}, we further investigate the alignment-dispersion characteristics of the learned representations.
\wx{Encouraging the cross-view alignment enables the view-aware representations to learn from each other and achieve mutual enhancement;
meanwhile, enlarging the cross-view dispersion between different users/bundles is excel at enhancing the representations' discriminative power.
Such a powerful representation learning comes with minimal space complexity and low time complexity.} Our main contributions are as follows:


\begin{itemize}[leftmargin=*]
    \item To the best of our knowledge, we are \za{among} the first to \za{formulate} the cross-view cooperative association in bundle recommendation, \za{providing a new research line worthy of further exploration.}
    \item We propose a simple yet effective bundle recommender, CrossCBR, to model the cooperative association between two views via cross-view contrastive learning.
    \item Our model outperforms SOTA baselines by a large margin on three public datasets, while requires largely reduced training time. 
    We also demonstrate how the idea of CrossCBR can be generalized to a broader scope of tasks.
\end{itemize}

\section{Methodology} \label{sec:methodology}
\begin{figure*}
    \centering
    \includegraphics[width = 0.9\linewidth]{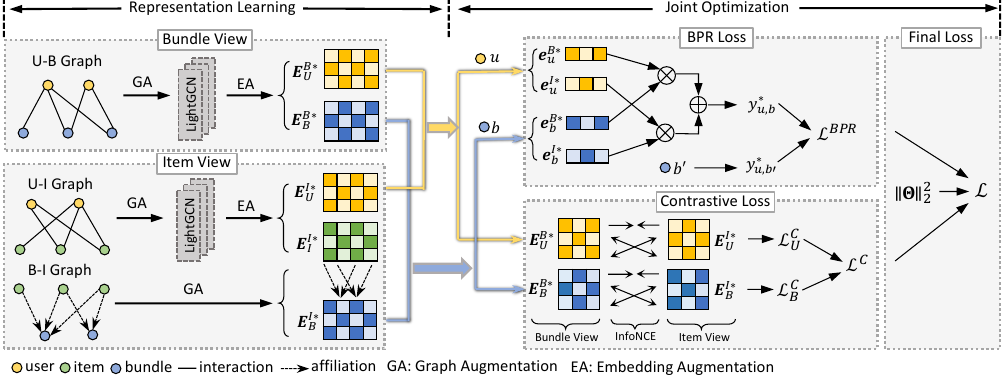}
    \vspace{-0.1in}
    \caption{The overall framework of CrossCBR consists of two parts: (1) representation learning for the two views of users and bundles and (2) the joint optimization of the BPR loss $\mathcal{L}^{BPR}$ and contrastive loss $\mathcal{L}^C$.}
    \vspace{-0.15in}
    \label{fig:framework}
\end{figure*}

In this section, we first formulate the task of bundle recommendation and the present our CrossCBR, as shown in Figure~\ref{fig:framework}.
The in-depth discussion of the working mechanism and analysis of computational complexity for CrossCBR are followed.

\subsection{Problem Formulation} \label{subsec:problem_formulation}
Given a set of users $\mathcal{U}=\{u_1, u_2, \cdots, u_M\}$, a set of bundles $\mathcal{B}=\{b_1, b_2, \cdots, b_N\}$, and a set of items $\mathcal{I}=\{i_1, i_2, \cdots, i_O\}$, where $M$, $N$, and $O$ are the number of users, bundles, and items, respectively. The user-bundle interactions, user-item interactions, and bundle-item affiliations are denoted as $\mathbf{X}_{M \times N}=\{x_{ub}|u\in{\mathcal{U}},b\in{\mathcal{B}}\}$, $\mathbf{Y}_{M \times O}=\{y_{ui}|u\in{\mathcal{U}},i\in{\mathcal{I}}\}$, and $\mathbf{Z}_{N \times O}=\{z_{bi}|b\in{\mathcal{B}},i\in{\mathcal{I}}\}$, respectively. $x_{ub}, y_{ui}, z_{bi} \in \{0, 1\}$, where $1$ \za{represents} an interaction between the user-bundle or user-item pair, or the item belongs to a certain bundle. 
Note that since we deduplicate the historical bundle and item interactions for each user, 
each element of $X$ and $Y$ is a binary value \za{rather than} an integer. In addition, $X$ and $Y$ are separately generated, where users are allowed to directly interact with both bundles and individual items. Therefore, $X$ and $Y$ contain different information, which heuristically enables the cooperative effect between the two different views. 
The goal of bundle recommendation task is to learn a model from the historical $\{X, Y, O\}$ and predict the unseen user-bundle interactions in $X$.

\subsection{Learning of Two Views' Representations} \label{subsec:representation_learning}
For the first component of CrossCBR, we aim to learn the representations \za{from} the two views: bundle and item view. 
Despite the \za{effectiveness of} two views' representation learning module of BGCN~\cite{chang2020bundle}, its \za{partial} designs of graph construction and graph learning are useless or even harmful~\cite{he2020lightgcn}, especially under the circumstance of utilizing the contrastive learning. 
\za{Here} we \za{devise} our simpler yet more effective representation learning approach.

\subsubsection{Bundle-view Representation Learning}
In order to learn the user and bundle representations \za{from} the bundle view, we first construct a user-bundle bipartite graph, \ie U-B graph, based on the user-bundle interaction matrix $\mathbf{X}$. We \za{then} employ the prevailing GNN-based recommendation framework LightGCN~\cite{he2020lightgcn} to learn the representations of both user and bundle. Specifically, we conduct information propagation over the U-B graph, and the $k$-th layer's information propagation is denoted as:
\begin{equation} \label{eq_2}
\left\{
\begin{aligned}
    \mathbf{e}_{u}^{B(k)}=\sum_{b \in \mathcal{N}_u}{\frac{1}{\sqrt{|\mathcal{N}_u|}\sqrt{|\mathcal{N}_b|}}\mathbf{e}^{B(k-1)}_{b}}, \\
    \mathbf{e}_{b}^{B(k)}=\sum_{u \in \mathcal{N}_b}{\frac{1}{\sqrt{|\mathcal{N}_b|}\sqrt{|\mathcal{N}_u|}}\mathbf{e}^{B(k-1)}_{u}},
\end{aligned}
\right.
\end{equation}
where $\mathbf{e}_{u}^{B(k)}, \mathbf{e}_{b}^{B(k)} \in \mathbb{R}^{d}$ are the $k$-th layer's information propagated to user $u$ and bundle $b$; $d$ is the embedding dimensionality; the superscript $B$ indicates the bundle view; $\mathbf{e}_{u}^{B(0)}$ and $\mathbf{e}_{b}^{B(0)}$ are randomly initialized at the beginning of the training; $\mathcal{N}_u$ and $\mathcal{N}_b$ are the neighbors of the user $u$ and bundle $b$ in the U-B graph. 

We follow LightGCN to remove the self-connections from the U-B graph and the nonlinear transformation from the information propagation function. We will empirically demonstrate that such simplifications, which BGCN does not take into account, are truly helpful for better performance \za{(\cf Section \ref{subsec:ablation_study}) }. 
More importantly, we do not incorporate the bundle-bundle connections, which are introduced by BGCN and calculated from the degree of overlapped items between the two bundles. 
The reason lies in \za{the fact} that bundle-bundle overlap information can be distilled from the item view through the cross-view contrastive learning (\za{\cf Section \ref{subsec:contrastive_loss}}). Meanwhile, the removal of extra bundle-bundle connections can further reduce the computational costs during the graph learning. 

We \za{concatenate} all $K$ layers' \za{embedding to combine the information received from neighbors of different depths}. 
The final bundle-view representations $\mathbf{e}^{\za{B}*}_{u}$ and \za{$\mathbf{e}^{B*}_b$} are denoted as:
\begin{equation} \label{eq_3}
    \mathbf{e}^{B*}_{u} = \sum^K_{k=0}{\mathbf{e}^{B(k)}_{u}}, \ \ \ \ \mathbf{e}^{B*}_b = \sum^K_{k=0}{\mathbf{e}^{B(k)}_b}.
\end{equation}

\subsubsection{Item-view Representation Learning}
In order to learn the user and bundle representations \za{from} the item view, we first build two bipartite graphs, \ie U-I and B-I graph, according to the user-item interactions $\mathbf{Y}$ and bundle-item affiliations \za{$\mathbf{Z}$}, respectively. 
Similar \za{to} the U-B graph learning, we learn user and item representations using LightGCN. 
The obtained user representations are the item-view user representations, and the item-view bundle representations are obtained by performing average pooling over the item-view item representations guided by the B-I graph. Specifically, the information propagation over the U-I graph is \za{defined} as:
\begin{equation} \label{eq_4}
\left\{
\begin{aligned}
    \mathbf{e}_{u}^{I(k)}=\sum_{i \in \mathcal{N}_u}{\frac{1}{\sqrt{|\mathcal{N}_u|}\sqrt{|\mathcal{N}_i|}}\mathbf{e}^{I(k-1)}_i}, \\
    \mathbf{e}_{i}^{I(k)}=\sum_{u \in \mathcal{N}_i}{\frac{1}{\sqrt{|\mathcal{N}_i|}\sqrt{|\mathcal{N}_u|}}\mathbf{e}^{I(k-1)}_{u}},
\end{aligned}
\right.
\end{equation}
where $\mathbf{e}_{u}^{I(k)}, \mathbf{e}_{i}^{I(k)} \in \mathbb{R}^{d}$ are the $k$-th layer's information propagated to user $u$ and item $i$, respectively; the superscript $I$ \za{refers to} the item view; $\mathbf{e}_{i}^{I(0)}$ \za{is} randomly initialized; $\mathcal{N}_u$ and $\mathcal{N}_i$ are the neighbors of the user $u$ and item $i$ in the U-I graph. 
We follow BGCN and share the parameters of $\mathbf{e}_{u}^{I(0)}$ with $\mathbf{e}_{u}^{B(0)}$, which empirically does not affect the performance but largely reduces the \za{number of} parameters. 
Meanwhile, such initial layer's parameters sharing between two views is too weak even impossible to capture the cross-view cooperative association (\cf CrossCBR-CL in Section \ref{subsec:ablation_study}). 
Similar \za{to} U-B graph, we also remove the self-connections from the U-I graph and nonlinear feature transformation from the information propagation function. 
And a layer aggregation operation is \za{adopted} after $K$ layers of information propagation, formulated as follows:
\begin{equation} \label{eq_5}
    \mathbf{e}^{I*}_{u} = \sum^K_{k=0}{\mathbf{e}^{I(k)}_{u}}, \ \ \ \ \mathbf{e}^{I*}_{i} = \sum^K_{k=0}{\mathbf{e}^{I(k)}_{i}},
\end{equation}
where $\mathbf{e}^{I*}_{u}$ and $\mathbf{e}^{I*}_{i}$ are the item-view user and item representations, respectively. Based on the item-view item representation and the B-I graph, we can obtain the item-view bundle representations $\mathbf{e}^{I*}_{b}$ through average pooling, denoted as:
\begin{equation} \label{eq_6}
    \mathbf{e}^{I*}_{b} = \frac{1}{|\mathcal{N}_b|}\sum_{i \in \mathcal{N}_b}\mathbf{e}^{I*}_i,
\end{equation}
where $\mathcal{N}_b$ represents the \za{set of items} a certain bundle $b$ contains. 

In summary, we can \za{learn} the representations of all users and bundles \za{from} two views, denoted as $\mathbf{E}_U^{B*}, \mathbf{E}_U^{I*} \in \mathbb{R}^{M \times d}$ and $\mathbf{E}_B^{B*}, \mathbf{E}_B^{I*} \in \mathbb{R}^{N \times d}$, where the superscripts $B$ and $I$ \za{stand for} the bundle and item view, respectively; and the subscripts $U$ and $B$ \za{indicate} the whole user and bundle set, respectively ($\mathbf{E}_I^{I*} \in \mathbb{R}^{O \times d}$ are the representations of all items in the item view). Thereafter, given a user $u$ and a bundle $b$, we can obtain their bundle-view representations, \ie $\mathbf{e}^{B*}_u$ and $\mathbf{e}^{B*}_b$, and their item-view representations, \ie $\mathbf{e}^{I*}_u$ and $\mathbf{e}^{I*}_b$. 

\subsection{Cross-view Contrastive Learning} \label{subsec:contrastive_learning}
We \za{devise the critical} component to model the cross-view cooperative association \za{via contrastive learning}. We first \za{present} the data augmentation methods, followed by the contrastive loss. 

\subsubsection{Data Augmentation}
The \za{main} idea of \za{self-supervised} contrastive learning is to \za{encourage} the representation affinity among various views of the same object, while \za{at the same time} enlarge the representation dispersion of different objects \cite{wang2020understanding}. 
In practice, if multiple views naturally exist for each object, \eg images taken from different angle\za{s}, or the bundle and item view in bundle recommendation, the contrastive loss can be directly applied. 
On the other hand, in many scenarios, multiple views are not available, and data augmentation is leveraged to generate multiple views from the original data~\cite{chen2020simple,gao2021simcse,wu2021self}. 
Proper data augmentation not only release the (multi-view) data constraint for applying contrastive learning, but also may \za{improve} the robustness to counter potential noise. Therefore, while keeping the original preservation (no augmentation) as the default setting, we also introduce two simple data augmentation methods: graph- and embedding-based augmentations.

\textbf{Graph-based Augmentation}. The main purpose of graph-based augmentation is to generate augmented data by revising the graph structure~\cite{wu2021self}. We adopt a simple random augmentation method of edge dropout (ED), which randomly removes a certain proportion (dropout ratio $\rho$) of edges from the original graph. The rationale behind edge dropout lies in that the core local structure of the graph is preserved. Therefore, the robustness of learned representations may be enhanced to counter certain noise.


\textbf{Embedding-based Augmentation}. Different from the graph-based augmentation, which can be applied only to graph data, embedding-based augmentations are more general and suitable for any deep representation learning based methods~\cite{gao2021simcse}. The \za{major} idea is to \za{vary} the learned representation embeddings regardless of how the embeddings are obtained. We employ message dropout (MD), which randomly masks some elements of the propagated embeddings with a certain dropout ratio $\rho$ during the graph learning. 


\textbf{Original Preservation}. We name the approach without any data augmentation as original preservation (OP), where no randomness is introduced and only the original representations are preserved. Since the two views in bundle recommendation are obtained from different sources of data, their representations are distinctive sufficiently to work well.

To avoid the abuse of notations, after the data augmentation, we still use the same notations of $\mathbf{e}^{B*}_u, \mathbf{e}^{B*}_b, \mathbf{e}^{I*}_u, \mathbf{e}^{I*}_b$ to denote the embeddings for the bundle-view user, bundle-view bundle, item-view user, and item-view bundle, respectively.

\subsubsection{Cross-view Contrastive Loss} \label{subsec:contrastive_loss}
We \za{leverage} the cross-view contrastive loss \za{to optimize} two-view representations. As the motivations \za{illustrate} in Figure~\ref{fig:motivation}, each view captures a distinctive aspect of user's preference, and the two views have to work cooperatively to maximize the overall modeling capacity. 
To model the cross-view cooperative association, we employ the cross-view contrastive loss (we leave other potential modeling solutions for future work). We adopt the popular InfoNCE~\cite{gutmann2010noise} loss built upon the cross-view representations of users and bundles, respectively. 
More precisely, the contrastive loss \za{is able to simultaneously encourage the alignment of the same user/bundle from different views and enforce the separation of different users/bundles.}
The equations are as follows:
\begin{equation} \label{eq:user_contrastive_loss}
    \mathcal{L}^C_{U} = \frac{1}{|\mathcal{U}|} \sum_{u \in \mathcal{U}}{-\text{log}\frac
    {\text{exp}({s(\mathbf{e}^{B*}_u, \mathbf{e}^{I*}_u)/\tau})}
    {\sum_{v \in \mathcal{U}}{\text{exp}({s(\mathbf{e}^{B*}_u, \mathbf{e}^{I*}_v)/\tau})}}},
\end{equation}
\begin{equation} \label{eq:bundle_contrastive_loss}
    \mathcal{L}^C_{B} = \frac{1}{|\mathcal{B}|} \sum_{b \in \mathcal{B}}{-\text{log}\frac
    {\text{exp}({s(\mathbf{e}^{B*}_b, \mathbf{e}^{I*}_b)/\tau})}
    {\sum_{p \in \mathcal{B}}{\text{exp}({s(\mathbf{e}^{B*}_b, \mathbf{e}^{I*}_p)/\tau})}}},
\end{equation}
where $\mathcal{L}^C_U$ and $\mathcal{L}^C_B$ denote the cross-view contrastive losses for users and bundles, respectively; $s(\cdot,\cdot)$ is the cosine similarity function; $\tau$ is a hyper-parameter known as the \textit{temperature}. We follow SGL \cite{wu2021self} to perform in-batch negative sampling to construct the negative pairs. By averaging the two cross-view contrastive losses, we obtain the final contastive loss $\mathcal{L}^C$:
\begin{equation} \label{eq:final_contrastive_loss}
    \mathcal{L}^C = \frac{1}{2}(\mathcal{L}^C_U + \mathcal{L}^C_B).
\end{equation}

\subsection{Prediction and Optimization}
To obtain the final prediction for recommendation, we first utilize the inner-product to calculate the item-view and bundle-view predictions, and additively combine them \za{for} the final prediction.
\begin{equation} \label{eq:prediction}
    y^*_{u,b} = {\mathbf{e}^{B*}_u}^{\intercal}{\mathbf{e}^{B*}_b} + {\mathbf{e}^{I*}_u}^{\intercal}{\mathbf{e}^{I*}_b}.
\end{equation}
The \za{conventional} Bayesian Personalized Ranking (BPR) loss \cite{rendle2012bpr} \za{is used} as the main loss.
\begin{equation} \label{eq:bpr_loss}
    \mathcal{L}^{BPR} = \sum_{(u,b,b^{\prime}) \in Q}{-\text{ln} \sigma (y^*_{u,b} - y^*_{u,b^{\prime}})}.
\end{equation}
where $Q = \{(u,b,b^{\prime})|u \in \mathcal{U}, b, b^{\prime} \in \mathcal{B}, x_{ub}=1, x_{ub^{\prime}}=0 \}$, $\sigma(\cdot)$ is the sigmoid function. We achieve the final loss $\mathcal{L}$ by weighted combing the BPR loss $\mathcal{L}^{BPR}$, the contrastive loss $\mathcal{L}^C$, and the L2 regularization term ${\Vert \mathbf{\Theta} \rVert}_2^2$:
\begin{equation} \label{eq:total_loss}
    \mathcal{L} = \mathcal{L}^{BPR} + \lambda_{1}{\mathcal{L}^C} + \lambda_{2}{\Vert \mathbf{\Theta} \rVert}_2^2,
\end{equation}
where $\lambda_1$ and $\lambda_2$ are the \za{hyperparameters} to balance the three terms, and $\mathbf{\Theta} = \{\mathbf{E}_{U}^{B(0)},\mathbf{E}_{B}^{B(0)},\mathbf{E}_{I}^{I(0)}\}$ are all of the model parameters.

\subsection{Model Discussion} \label{subsec:model_discussion}
\za{Integrating the cross-view contrastive loss into BPR loss can provide an additional regularization for representation learning \cite{zhang2021unleashing}.}
\za{Such an effect encourages the bundle recommender to achieve} the cross-view alignment of the same user/bundle \za{and impose the} dispersion of \za{different user/bundle pairs in both ego-view and cross-view.}

By \za{enforcing} the cross-view alignment of the same user/bundle, the distinctive information contained in each view will be distilled to the other view. 
Therefore, both views' representations can be enhanced.
Consequently, the combined prediction of the two views \za{can be} further \za{boosted}. 
\za{See more details and results in Section~\ref{subsubsec:mutual-enhancement}.}

\begin{figure}
    \centering
    \includegraphics[width = 0.8\linewidth]{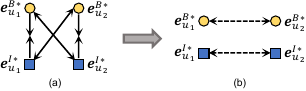}
    \vspace{-0.1in}
    \caption{The illustration of the direct (a) and indirect (b) effects of the cross-view contrastive loss.}
    \label{fig:CL_analysis}
    \vspace{-0.2in}
\end{figure}

In addition to the cross-view alignment, the effect of representation dispersion is also \za{pivotal}. 
Based on Equations~\ref{eq:user_contrastive_loss} and \ref{eq:bundle_contrastive_loss}, it seems only the item/bundle pairs across different views are pushed away. 
However, the cross-view \za{alignment also} acts as a bridge to make the \za{distinct} user/bundle pairs within the same view be \za{widely separated}. 
Let's take a pair of users $u_1$ and $u_2$ as \za{an} example, shown in Figure~\ref{fig:CL_analysis}. 
The direct effect of the contrastive loss is to pull close the pairs $(\mathbf{e}^{B*}_{u_1}, \mathbf{e}^{I*}_{u_1})$ and $(\mathbf{e}^{B*}_{u_2}, \mathbf{e}^{I*}_{u_2})$ while push \za{apart} the pairs $(\mathbf{e}^{B*}_{u_1}, \mathbf{e}^{I*}_{u_2})$ and $(\mathbf{e}^{I*}_{u_1}, \mathbf{e}^{B*}_{u_2})$. 
Consequently, as an indirect effect, the \za{distance between} the representations of $u_1$ and $u_2$ in the same view (\ie $(\mathbf{e}^{B*}_{u_1}, \mathbf{e}^{B*}_{u_2})$ and $(\mathbf{e}^{I*}_{u_1}, \mathbf{e}^{I*}_{u_2})$) are also \za{enlarged}.
Therefore, our proposed cross-view contrastive loss can enhance the discriminative capability of representations in both ego-view and cross-view, resulting in better \za{bundle} recommendation \za{quality}. To be noted, solely enlarging the cross-view dispersion without encouraging the cross-view alignment cannot achieve the effect of ego-view dispersion, thus cannot enhance the self-discrimination of the representations.
We will justify this effect by analyzing the alignment-dispersion characteristics of representations in Section~\ref{subsubsec:alignment-dispersion}.


\subsection{Complexity Analysis}
In terms of space complexity, the parameters of CrossCBR are minimal and only include three sets of embeddings: $\mathbf{E}^{B(0)}_U$, $\mathbf{E}^{B(0)}_{B}$, and $\mathbf{E}^{I(0)}_I$. Therefore, the space complexity of CrossCBR is $\mathcal{O}((M+N+O)d)$. Our model is more concise than BGCN due to the removal of the feature transformation matrices.

In terms of time complexity, the main computational cost lies in the two views' graph learning and the cross-view contrastive loss. Note we just focus on the main setting of original preservation. The time complexity of graph learning is $O(|E|Kds\frac{|E|}{T})$, where $|E|$ is the number of all edges in U-B and U-I graphs, $K$ is the number of propagation layers, $d$ is the embedding size, $s$ is the number of epochs, $T$ is the batch size. For comparison, the time complexity of BGCN is $O((|E|d+|V|d^2)Ks\frac{|E|}{T})$, where $|V|$ is the number of nodes in U-B and U-I graphs. And the time complexity of graph learning part in CrossCBR is smaller than that of BGCN due to the removal of the feature transformation layers and smaller $|E|$ (due to the removal of self-connections and bundle-bundle connections). The time complexity of calculating the contrastive loss during training is $O(2d(|E|+T^{2})s\frac{|E|}{T})$. Even though the matrix multiplication of contrastive loss has cubic complexity ($O(dT^{2})$), the time used in practice is very limited due to the acceleration of both hardware GPU and the optimized libraries, justified by the experiments in Section~\ref{subsubsec:computational-efficiency}. To be noted, the inference of CrossCBR has the identical time complexity with that of BGCN.
\section{Experiments} \label{sec:experiment}
To evaluate our proposed approach, we conduct experiments on three public datasets: Youshu, NetEase, and iFashion. In particular, we aim to answer the following research questions:

\begin{itemize}[leftmargin=*]
    \item \textbf{RQ1: } Can CrossCBR outperform the SOTA baseline models?
    \item \textbf{RQ2: } Are all the key components effective \wrt performance?
    \item \textbf{RQ3: } Whether the cross-view contrastive learning works as we expected, \ie achieving cross-view mutual enhancement and alignment and dispersion in the representation space?
    \item \textbf{RQ4: } What about the hyper-parameter sensitivity and training efficiency of the model?
\end{itemize}

\subsection{Experimental Settings}
We follow the previous works~\cite{deng2020personalized,chang2020bundle} to adopt the two established bundle recommendation datasets: Youshu~\cite{chen2019matching} for book list recommendation and NetEase~\cite{cao2017embedding} for music playlist recommendation. In addition, we introduce another online fashion outfit recommendation dataset iFashion~\cite{chen2019pog}, where the outfit consisted of individual fashion items is treated as bundle. We follow the outfit recommendation setting~\cite{li2020hierarchical} to preprocess the iFashion dataset by the 20-core rule for users and 10-core rule for outfits. All the three datasets have all the required data, \ie user-bundle interactions, user-item interactions, and bundle-item affiliation. The statistics of the datasets are listed in Table~\ref{tab:dataset}. To be noted, the three datasets are diverse \wrt both application scenarios and the statistical characteristics (various scales of interactions and bundle sizes), ensuring the model's robustness to such variance. The training/validation/testing sets are randomly split with the ratio of 70\%/10\%/20\%. Recall@K and NDCG@K are used as the evaluation metrics, where K $\in$ \{20, 40\}. And NDCG@20 is used to select the best model based on the validation set, and all the items are ranked during testing~\cite{wang2019neural}.

\begin{table}[t]
\begin{center}
\caption{Dataset Statistics.}
\label{tab:dataset}
\vspace{-0.1in}
\resizebox{0.45\textwidth}{!}{
    \begin{tabular}{ccccccc}
        \hline
        Dataset & \#U & \#I & \#B & \#U-I & \#U-B & \#Avg.I/B \\
        \hline
        Youshu   & 8,039  & 32,770  & 4,771  & 138,515   & 51,377    & 37.03 \\
        NetEase  & 18,528 & 123,628 & 22,864 & 1,128,065 & 302,303   & 77.80 \\
        iFashion & 53,897 & 42,563 & 27,694 & 2,290,645 & 1,679,708 & 3.86 \\
        \hline
    \end{tabular}
}
\end{center}
\vspace{-0.15in}
\end{table}

\begin{table*}[t]
\caption{The overall performance comparison, where Rec is short of Recall. \za{Note that the improvement achieved by CrossCBR is significant ($p$-value $\ll 0.05$).}}
\vspace{-0.1in}
\label{tab:overall_performance}
\centering
\setlength{\tabcolsep}{1mm}{
    \resizebox{\textwidth}{!}{
        \begin{tabular}{l | cccc | cccc | cccc}
        \hline
        \multirow{2}{*}{Model} & \multicolumn{4}{c|}{Youshu} &\multicolumn{4}{c|}{NetEase} &\multicolumn{4}{c}{iFashion} \\
        \cline{2-13} & Rec@20 & NDCG@20 & Rec@40 & NDCG@40 & Rec@20 & NDCG@20 & Rec@40 & NDCG@40 & Rec@20 & NDCG@20 & Rec@40 & NDCG@40 \\
        \hline
        \hline
        \textbf{MFBPR}     & 0.1959	& 0.1117 & 0.2735 & 0.1320 & 0.0355 & 0.0181 & 0.0600 & 0.0246 & 0.0752 & 0.0542 & 0.1162 & 0.0687 \\
        \textbf{LightGCN}  & 0.2286 & 0.1344 & 0.3190 & 0.1592 & 0.0496 & 0.0254 & 0.0795 & 0.0334 & 0.0837 & 0.0612 & 0.1284 & 0.0770 \\ 
        \textbf{SGL} & \underline{0.2568} & \underline{0.1527} & \underline{0.3537} & \underline{0.1790} & \underline{0.0687} & \underline{0.0368} & \underline{0.1058} & \underline{0.0467} & \underline{0.0933} & \underline{0.0690} & \underline{0.1389} & \underline{0.0851} \\
        \hline
        \hline
        \textbf{DAM} & 0.2082 & 0.1198 & 0.2890 & 0.1418 & 0.0411 & 0.0210 & 0.0690 & 0.0281 & 0.0629 & 0.0450 & 0.0995 & 0.0579 \\
        \textbf{BundleNet} & 0.1895 & 0.1125 & 0.2675 & 0.1335 & 0.0391 & 0.0201 & 0.0661 & 0.0271 & 0.0626 & 0.0447 & 0.0986 & 0.0574 \\
        \textbf{BGCN}      & 0.2347 & 0.1345 & 0.3248 & 0.1593 & 0.0491 & 0.0258 & 0.0829 & 0.0346 & 0.0733 & 0.0531 & 0.1128 & 0.0671 \\
        \hline
        \hline 			 						
        \textbf{CrossCBR} & \textbf{0.2813} & \textbf{0.1668} & \textbf{0.3785} & \textbf{0.1938} & \textbf{0.0842} & \textbf{0.0457} & \textbf{0.1264} & \textbf{0.0569} & \textbf{0.1173} & \textbf{0.0895} & \textbf{0.1699} & \textbf{0.1080} \\				
        \textbf{\%Improv.} & 9.57 & 9.26 & 7.02 & 8.28 & 22.57 & 24.33 & 19.48 & 21.96 & 25.76 & 29.63 & 22.33 & 26.85 \\
        \hline
        \end{tabular}
    }
}
\vspace{-0.1in}
\end{table*}

\subsubsection{Compared Methods}
In terms of baselines, we select both general user-item recommendation models and bundle-specific recommendation models to compare with our proposed method.

\noindent \textbf{The User-item Recommendation Models} treat the bundle as a special type of \textit{item}, only using the user-bundle interactions without considering the affiliated items within the bundle. We select the following SOTA methods: (1) \textbf{MFBPR}~\cite{rendle2012bpr}: Matrix Factorization optimized with the Bayesian Personalized Ranking (BPR) loss; (2) \textbf{LightGCN}~\cite{he2020lightgcn}: a GNN- and CF-based recommendation model, which utilizes a light-version graph learning kernel; and (3) \textbf{SGL}~\cite{wu2021self}: it enhances the LightGCN model with contrastive graph learning and achieves SOTA performance.

\noindent \textbf{The Bundle-specific Recommendation Models} are designed for bundle recommendation and utilize all the user-bundle interactions, user-item interactions, and bundle-item affiliation data. We consider the following models: (1) \textbf{DAM}~\cite{chen2019matching}: it uses an attention mechanism to learn bundle representations over the affiliated items and employs multi-task learning to optimize both user-item and user-bundle interactions; (2) \textbf{BundleNet}~\cite{deng2020personalized}: it builds a user-bundle-item tripartite graph, leverages GCN to learn the representations, and applies multi-task learning; and (3) \textbf{BGCN}~\cite{chang2020bundle,chang2021bundle}: it decomposes the user-bundle-item relations into two separate views, builds two graphs (\ie bundle-view graph and item-view graph), uses GCN to learn representations, makes prediction by summing the two views' predictions, and achieves SOTA performance. There are also some earlier works (\eg \cite{cao2017embedding}) that have been turned to be inferior to the methods listed above, and we do not consider them.

\subsubsection{Hyper-parameter Settings}
For all methods, the embedding size is set as 64, Xavier normal initialization~\cite{glorot2010understanding} is adopted, the models are optimized using Adam optimizer~\cite{kingma2014adam} with the learning rate 0.001, and the batch size is set as 2048. For our method, we tune the hyper-parameters $K$, $\lambda_1$, $\lambda_2$, $\tau$, and $\rho$ with the ranges of $\{1, 2, 3\}$, $\{0.01, 0.04, 0.1, 0.5, 1\}$, $\{10^{-6}, 10^{-5}, 2 \times 10^{-5}, 4 \times 10^{-5}, 10^{-4} \}$, $\{0.05, \\ 0.1, 0.15, 0.2, 0.25, 0.3, 0.4, 0.5 \}$, and $\{0, 0.1, 0.2, 0.5\}$. For graph augmentation, we follow SGL to drop edges by every epoch. For baseline methods, we adopt the results of MFBPR, DAM, and BGCN on Youshu and NetEase datasets based on those reported in \cite{chang2020bundle}, since their settings are the same with ours. We implement all the other baselines by ourselves. All the models are trained using Pytorch 1.9.0, NVIDIA Titan-XP and Titan-V GPUs.

\subsection{Performance Comparison (RQ1)} \label{subsec:performance_comparison}
We first compare the overall recommendation performance of CrossCBR with \za{both user-item recommendation baselines and bundle-specific recommendation baselines} on three datasets, as shown in Table~\ref{tab:overall_performance}.
\za{The best performing methods are bold, while the strongest baselines are underlined; \%Improv. measures the relative improvements of CrossCBR over the strongest baselines. We observe that:}

In terms of the general user-item recommendation models, LightGCN \za{consistently} outperforms MFBPR, indicating the GNN-based method especially the LightGCN graph learning module is effective in modeling the user-bundle CF signals. 
SGL further improves the performance of LightGCN, demonstrating the great power of \za{contrastive loss} on the user-bundle bipartite graph. 
\za{Surprisingly}, SGL is the strongest baseline, which is even better than the strongest bundle-specific method (BGCN), implying the effectiveness of graph contrastive learning in recommendation. 
\za{Our proposed} CrossCBR performs better than SGL by a large margin, showing that the item view truly provides additional useful information and can enhance the \za{discirminative power of} the model.

\za{When considering} the bundle-specific models, BGCN performs best \za{among bundle-specific baselines, \ie DAM and BundleNet.}
\za{We attribute this success to} decomposing the users' preference into two views.
\za{Unfortunately}, BundleNet performs \za{poorly} since the user-bundle-item tripartite graph fails to differentiate behavioral similarity among users and content relatedness. 
CrossCBR \za{achieves significant gains over} all the bundle-specific baselines by a large margin, demonstrating the \za{effectiveness} of modeling the cross-view cooperative association. Our model performs consistently on all the three datasets that belong to varying application scenarios (including book, music, and fashion) and have different statistical characteristics (scales of interactions and bundle sizes). Therefore, out model turns out to be sufficiently robust.

\begin{table}[t]
\begin{center}
\caption{Ablation study of the key components of CrossCBR.}
\vspace{-0.1in}
\label{tab:ablation_study}
    \resizebox{0.45\textwidth}{!}{
        \begin{tabular}{l | cc | cc}
            \hline
            \multirow{2}{*}{Model} & \multicolumn{2}{c|}{NetEase} &\multicolumn{2}{c}{iFashion} \\
            \cline{2-5}
             & Rec@20 & NDCG@20 & Rec@20 & NDCG@20  \\
            \hline
            \textbf{BGCN} & 0.0491 & 0.0258 & 0.0733 & 0.0531 \\
            \hline
            \textbf{CrossCBR-CL} & 0.0608 & 0.0320 & 0.0852 & 0.0626 \\
            \textbf{CrossCBR\_A} & 0.0602 & 0.0314 & 0.0664 & 0.0474 \\
            \textbf{CrossCBR\_D} & 0.0410 & 0.0208 & 0.0568 & 0.0402 \\
            \hline
            \textbf{CrossCBR\_OP} & 0.0831 & 0.0452 & 0.1167 & 0.0891 \\
            \textbf{CrossCBR\_ED} & 0.0842 & 0.0457 & 0.1176 & 0.0891 \\
            \textbf{CrossCBR\_MD} & 0.0828 & 0.0457 & 0.1173 & 0.0895 \\
            \hline
            \textbf{CrossCBR+SC} & 0.0831 & 0.0455 & 0.1146 & 0.0873 \\
            \textbf{CrossCBR+BB} & 0.0828 & 0.0444 & 0.1163 & 0.0887 \\
            \hline
            \textbf{CrossCBR} & 0.0842 & 0.0457 & 0.1173 & 0.0895 \\
            \textbf{\%Improv.} & 71.43 & 77.29 & 52.50 & 64.54 \\
            \hline
        \end{tabular}
    }
\end{center}
\vspace{-0.15in}
\end{table}

\subsection{Ablation Study (RQ2)} \label{subsec:ablation_study}
To further evaluate the key innovative components of CrossCBR, we conduct a list of ablation studies as shown in Table~\ref{tab:ablation_study}, where the \%Improv. \za{quantifies} the relative improvement of CrossCBR over the SOTA bundle-specific model BGCN.

\subsubsection{Effectiveness of Cross-view Contrastive Learning}
To evaluate whether the cross-view contrastive learning \za{contributes} to the performance, we remove the contrastive loss $\mathcal{L}^C$ \za{in Equation \eqref{eq:final_contrastive_loss}} during training, \za{named CrossCBR-CL}. 
CrossCBR-CL \za{inevitably shows a severe} performance \za{decline} compared with CrossCBR, justifying the crucial role of \za{modelling} cross-view \za{information}. 
\za{Surprisingly}, CrossCBR-CL still \za{significantly} outperforms BGCN.
\za{We ascribe its success to utilizing the LightGCN kernel, which has been proved to be more effective than the typical GCN kernel used by BGCN \cite{he2020lightgcn}.}
To further identify the \za{characteristics} of alignment and dispersion, we implement CrossCBR\_A that only enlarges the cross-view alignment (using the negative cross-view cosine similarity to replace the contrastive loss), and CrossCBR\_D that only encourages the cross-view dispersion (setting the cosine similarity in numerator of the contrastive loss as static 1).
The results in Table~\ref{tab:ablation_study} demonstrates that the alignment and dispersion collaboratively contribute to the performance, while only modeling either of them can degrade the performance. Especially when only enlarging the cross-view dispersion (CrossCBR\_D), the model can hardly benefit from it or even collapse, justifying our discussion in Section~\ref{subsec:model_discussion}.


\subsubsection{Effectiveness of Data Augmentations}
We try various settings of data augmentations during the representation learning of the two views. CrossCBR\_OP corresponds to original preservation (\ie no augmentation), CrossCBR\_ED \za{represents} Edge Dropout of the graph-based augmentation method, and CrossCBR\_MD \za{refers} to Message Dropout of the embedding-based augmentation method. 
The results in Table~\ref{tab:ablation_study} demonstrate that the differences among the three data augmentation settings for CrossCBR are negligible compared with the performance gain over baselines. This phenomenon indicates that the distinction within the original data of the two views provides sufficient variance for the cross-view contrastive learning, while the variance introduced by random data augmentation is insignificant. More advanced and effective data augmentation methods can be explored in the future.

\subsubsection{The Impact of Simplification of Graph Structure}
CrossCBR simplifies the graph structure of BGCN by removing self-connections (SC) in the U-B and U-I graphs and the bundle-bundle (BB) connections in the U-B graph. To justify that these removals do not affect the performance, we specifically add SC and BB connections to our graph construction and obtain CrossCBR+SC and CrossCBR+BB, respectively. The results of both CrossCBR+SC and CrossCBR+BB indicate that both SC and BB contribute little or even none to CrossCBR. The reasons are two-fold. First, in terms of SC, the LightGCN kernel has no feature transformation and activation layers, resulting in the SC a simple summation of the node itself (layer 0), which adds no additional information. 
Second, for the BB connections, they are obtained by calculating the overlap degree of the two bundles according to the B-I graph. However, the BB overlap information can be distilled from the item view to the bundle view representations through the alignment effect of the cross-view contrastive loss. 

\subsection{Model Study (RQ3)} \label{subsec:model_study}
In this section, we conduct experiments to study: (1) whether the two views are mutually enhanced by the cross-view contrastive loss? and (2) whether the cross-view alignment and dispersion happen as we expected in the represenation space? 

\begin{figure}[t]
    \centering
    \includegraphics[width = 0.98\linewidth]{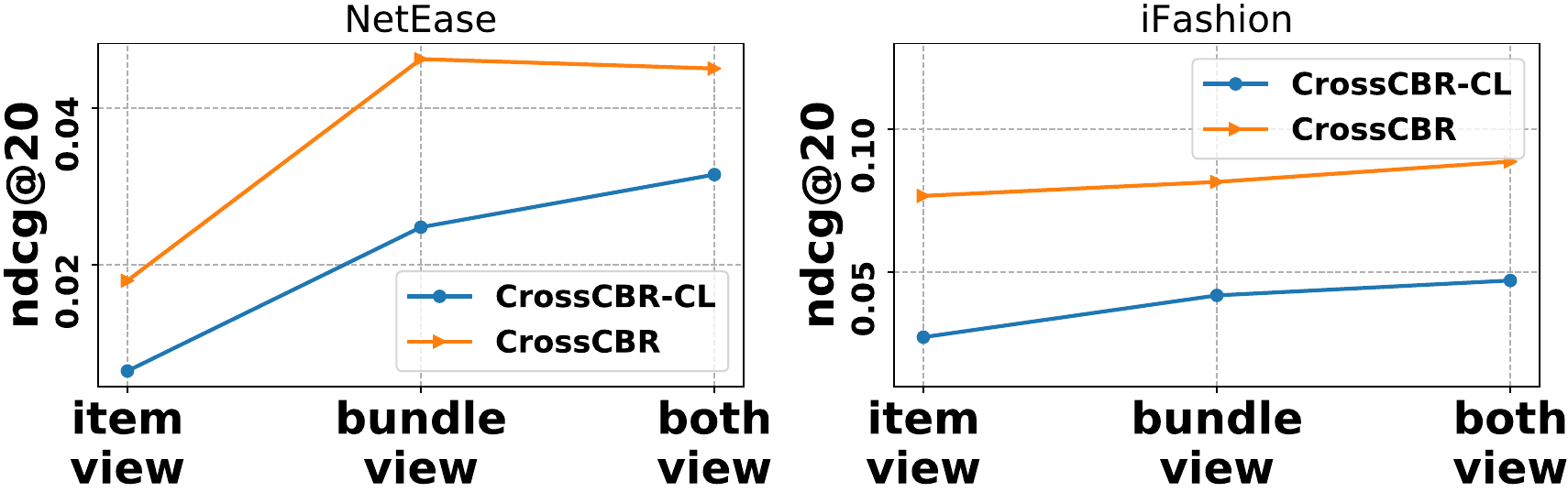}
    \vspace{-0.1in}
    \caption{The performance of CrossCBR and CrossCBR-CL \wrt predictions based on different views.}
    \label{fig:perf_views}
    \vspace{-0.15in}
\end{figure}

\subsubsection{Mutual Enhancement Effect} \label{subsubsec:mutual-enhancement}
To directly justify whether the cross-view contrastive learning can achieve cross-view mutual enhancement, we present the performance which is calculated solely based on ego-view predictions, \ie the bundle-view prediction uses $y_{u,b}^*={\mathbf{e}_u^{B*}}^{\intercal} \mathbf{e}_b^{B*}$, the item-view prediction uses $y_{u,b}^*={\mathbf{e}_u^{I*}}^{\intercal} \mathbf{e}_b^{I*}$, and the both-view prediction is identical with Equation~\ref{eq:prediction}. The results in Figure~\ref{fig:perf_views} indicate that using contrastive loss significantly improves the recommendation performance on all the three types of predictions. Interestingly, the bundle view prediction performs much better than that of the item view (even slightly better than the both-view prediction in NetEase), demonstrating that the bundle view plays a more significant role in bundle recommendation. This also helps explain why SGL, which is solely based on the user-bundle interactions, can outperform a list of bundle-specific methods. 

\begin{table}[t]
\begin{center}
\renewcommand{\arraystretch}{1.3}
\caption{The cross-view alignment and dispersion analysis of the representations. $\mathbf{A}$ denotes \textit{Alignment}; $\mathbf{D}$ denotes \textit{Dispersion}; superscripts $(C, B, I)$ denote the cross, bundle, and item view; subscripts $(U, B)$ stand for users and bundles.}
\vspace{-0.1in}
\label{tab:alignment_dispersion}
    \resizebox{0.45\textwidth}{!}{
        \begin{tabular}{c | cc | cc}
            \hline
            \multirow{2}{*}{Metrics} & \multicolumn{2}{c|}{NetEase} &\multicolumn{2}{c}{iFashion} \\
            \cline{2-5}
             & CrossCBR-CL & CrossCBR & CrossCBR-CL & CrossCBR  \\
            \hline
            $\mathbf{A}_U^C$ & 0.638 & 0.932 & 0.878 & 0.950 \\
            $\mathbf{D}_U^B$ & 0.313 & 0.049 & 0.331 & 0.014 \\
            $\mathbf{D}_U^I$ & 0.044 & 0.004 & 0.193 & 0.004 \\
            \hline
            $\mathbf{A}_B^C$ & 0.351 & 0.632 & 0.635 & 0.910 \\
            $\mathbf{D}_B^B$ & 0.040 & 0.042 & 0.052 & 0.033 \\
            $\mathbf{D}_B^I$ & 0.075 & 0.016 & 0.011 & 0.030 \\
            \hline
        \end{tabular}
    }
\end{center}
\vspace{-0.15in}
\end{table}

\subsubsection{Cross-View Alignment and Dispersion Analysis} \label{subsubsec:alignment-dispersion}
We analyze the cross-view alignment and dispersion characteristics of the representations regularized by the cross-view contrastive learning. Inspired by the alignment-uniformity analysis~\cite{wang2020understanding,gao2021simcse}, we adopt a simplified version to portray the cross-view alignment and dispersion of the representations. In particular, we just calculate the average cross-view \textit{cosine similarity} of the users (bundles) as the indication of alignment. Similarly, the average \textit{cosine similarity} between different users (bundles) of the same view (either item or bundle view) is the indication of the dispersion. Several interesting findings can be derived from the results shown in Table~\ref{tab:alignment_dispersion}. First, the cross-view alignment metrics of both user and bundle representations ($\mathbf{A}_U^C$ and $\mathbf{A}_B^C$) significantly increase after applying the contrastive loss, justifying that the cross-view contrastive loss can effectively pull the two views close to each other. Thereafter, each view can distill cooperative information from the other and they are mutually enhanced. Second, the dispersion of the user representations of both ego-views ($\mathbf{D}_U^B$ and $\mathbf{D}_U^I$) significantly reduces after applying the contrastive loss, verifying that the cross-view contrastive loss can improve the discrimination of the users in the ego-view ( \cf Section~\ref{subsec:model_discussion}). 
Third, the dispersion of the bundle representations ($\mathbf{D}_B^B$ and $\mathbf{D}_B^I$) does not consistently reduce like that of user's. It may because that the B-I graph directly determines the dispersion of item-view bundle representations via the pooling (\cf Equation~\ref{eq_6}), which is distilled to the bundle view. 

\subsection{Hyper-parameter and Computational Efficiency Analysis (RQ4)} \label{subsec:hyper-parameters}

\begin{figure}[t]
    \centering
    \includegraphics[width = 0.98\linewidth]{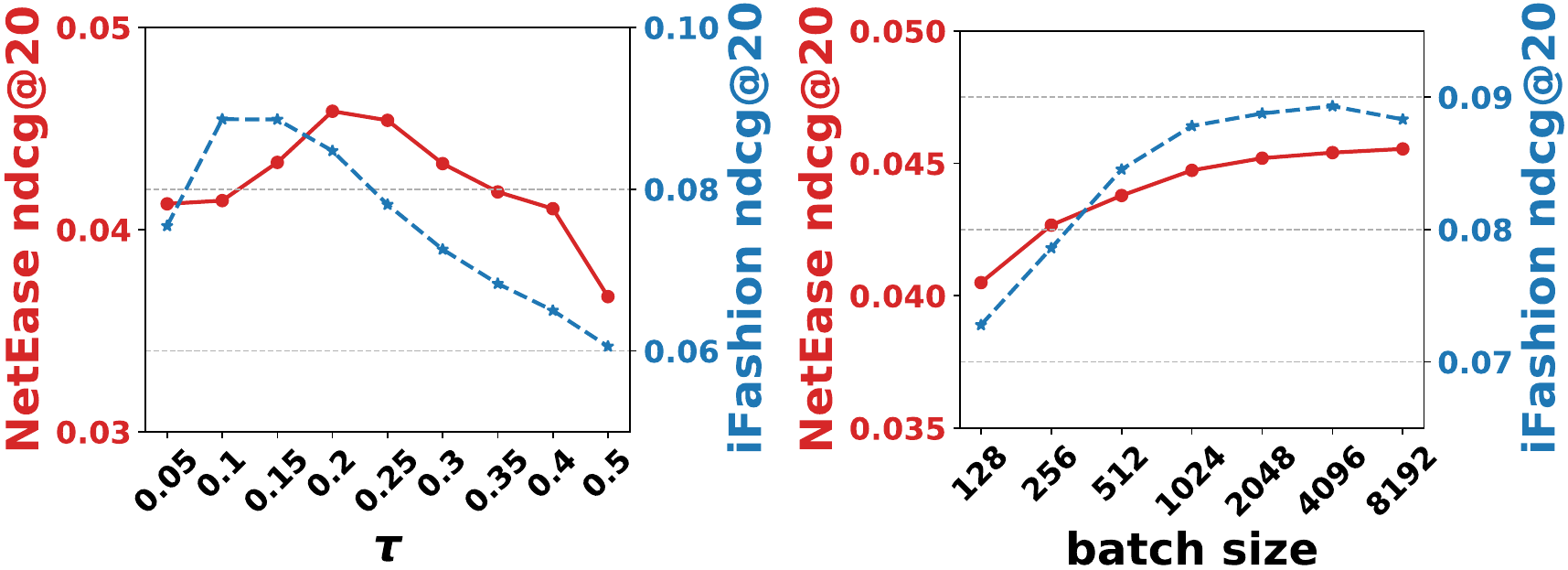}
    \vspace{-0.1in}
    \caption{The performance (NDCG@20) variance of CrossCBR \wrt the temperature $\tau$ and the batch size on both datasets of NetEase and iFashion.}
    \label{fig:perf_tau_bs}
    \vspace{-0.15in}
\end{figure}

\subsubsection{Hyper-parameter Analysis}
As shown in Figure~\ref{fig:perf_tau_bs}, CrossCBR is sensitive to the temperature $\tau$, and deviating from the best setting degrades the performance remarkably. 
To test how the batch size affects the performance, we gradually increase the batch size from 128 to 8192, and the performance first grows quickly and later reaches a plateau as shown in Figure~\ref{fig:perf_tau_bs}. 
We keep our default batch size as 2048, since it is widely adopted by the baselines' original implementation and performs sufficiently well in CrossCBR.

\subsubsection{Computational Efficiency} \label{subsubsec:computational-efficiency}
To evaluate the computational efficiency of our model, we compare the one-epoch training time \za{among the variants of} our model and BGCN.
We record ten consecutive training epochs and average them to obtain the one-epoch training time, as shown in Table~\ref{tab:training_time}~\footnote{The CPU is Intel(R) Xeon(R) CPU E5-2620 v3 with 12 cores, and the memory is 256GB. We use multiprocess dataloader to assure that the computation on CPU is not the bottleneck. In addition, the GPU utility is almost 100\% during the training of all the settings, therefore, the capacity of GPU dominates the overall training time.}. First of all, Cr\_OP is significantly more efficient than BGCN, demonstrating the efficiency of CrossCBR. Second, we compare Cr\_OP with several variants to further explicitly attribute the efficiency boost. In particular, Cr-CL approximates Cr\_OP, showing that the contrastive loss brings negligible computational costs.
Cr+SC and Cr+BB cost more training time than Cr\_OP, demonstrating that both SC and BB connections introduce extra costs during training. Especially on NetEase, the costs brought by the BB connections are about three times of Cr\_OP.

\begin{table}[t]
\begin{center}
\caption{The statistics of one-epoch training time (seconds) for CrossCBR and baselines on different devices, where the "Cr" is short of "CrossCBR".}
\vspace{-0.1in}
\label{tab:training_time}
    \resizebox{0.45\textwidth}{!}{
        \begin{tabular}{l |c|c|c|c|c|c}
            \hline
            Device & Dataset & BGCN & Cr-CL & \textbf{Cr\_OP} & Cr+SC & Cr+BB \\
            \hline
            \multirow{2}{*}{Titan XP} 
            & NetEase & 32.05 & 6.80 & 7.04 & 7.27 & 28.02 \\
            \cline{2-7}
            & iFashion & 63.47 & 46.05 & 46.74 & 47.61 & 56.42 \\ 
            \hline
            \multirow{2}{*}{Titan V} 
            & NetEase & 20.76 & 4.67 & 5.09 & 5.48 & 18.59 \\
            \cline{2-7}
            & iFashion & 38.76 & 29.48 & 30.02 & 30.31 & 35.01 \\
            \hline
        \end{tabular}
    }
\end{center}
\vspace{-0.2in}
\end{table}

\section{Related Work} \label{sec:related_work}
In this section, we briefly review the related works in two areas: (1) graph-based and bundle recommendation and (2) contrastive learning in recommendation.

\textbf{Graph-based and Bundle Recommendation}. Graph-based model has dominated the CF-based recommendation methods due to its superior capability in modeling the higher order interactions between users and items, especially the recent graph neural network-based methods~\cite{ying2018graph,wang2019neural,he2020lightgcn,chen2019semi}. Wang \etal propose NGCF~\cite{wang2019neural} to build a bipartite graph based on the user-item interaction matrix and employ graph convolutional network as the graph learning module. Following NGCF, He \etal propose to remove some redundant modules (\eg nonlinear feature transformation and activation function layers) from the NGCF model and significantly improve the performance, resulting in the LightGCN~\cite{he2020lightgcn} model. LightGCN has achieved great performance in various recommendation tasks~\cite{ding2021DGSR}, and our model is also base on this backbone.

Bundle recommendation aims to solve a special scenario of recommendation, \ie the recommended object is a bundle of items that related with a certain theme. Initial works just ignore the affiliated items of the bundle and just use an id to represent a bundle~\cite{rendle2010factorizing}. Following works recognize the importance of affiliated items and develop various models to capture the additional user-item interaction and bundle-item affiliation relations, such as EFM~\cite{cao2017embedding} and DAM~\cite{chen2019matching}. With the proliferation of GNN-based recommendation models, Deng \etal propose BundleNet~\cite{deng2020personalized} and Chang \etal propose BGCN~\cite{chang2020bundle,chang2021bundle}. However, BundleNet mixup the three types of relations among user, bundle, and item, while BGCN decompose the users' preference into item view and bundle view. The two-view representations effectively capture the two types of preferences, resulting in better performance. Our work is based on this two-view modeling framework, and we further emphasize the significance of the cooperative association modeling between the two views. Some related topics, such as set, basket, or package recommendation~\cite{hu2020modeling,qin2021world,li2021package} and bundle generation~\cite{bai2019personalized,chang2021bundle}, are different with our scenario in either the recommended object (a loosely/arbitrary co-occurred set/basket/package \textit{vs} a pre-defined bundle of items related with a theme) or the task (generation of bundle from items \textit{vs} recommending pre-defined bundles).

\textbf{Contrastive Learning in Recommendation}. Recently, contrastive learning regains popularity and achieves great success in computer vision~\cite{hjelm2018learning,oord2018representation,chen2020simple}, natural language processing~\cite{logeswaran2018efficient,gao2021simcse}, and graph learning~\cite{DIR,RGCL}. The community of recommender systems also seizes this trend and adapts contrastive learning into various recommendation tasks, such as general CF-based recommendation~\cite{wu2021self,zhou2021selfcf,zhou2021contrastive}, sequential and session recommendation~\cite{zhou2020s3,xie2020contrastive,liu2021contrastive,xia2020self,xia2021self}, multimedia and social recommendation~\cite{wei2021contrastive,yu2021socially}, \etc The key of introducing contrastive learning into recommender systems lies in proper construction of contrastive pairs. One branch of current approaches are based on various data augmentations to create more views from the original data. For example, SGL~\cite{chang2020bundle} adopts various graph augmentation methods (\eg dege dropout or random walk), and CL4SRec~\cite{xie2020contrastive} and CoSeRec~\cite{liu2021contrastive} apply different sequence augmentation methods (\eg insertion, deletioin, and reordering \etc). Another branch of methods aim at mining multiple views that exist in the data. For example, COTREC~\cite{xia2021self} builds two views (\ie an item view and a session view) to learn the session representations from two sources of data (\ie item transition graph of a session and session-session similarity graph), and apply contrastive learning based on the two views. CLCRec~\cite{wei2021contrastive} treats different modalities and the user/item as different views to build contrastive pairs. In this work, we unify both types of methods: build two views from different data sources and apply data augmentations.

\section{Conclusion and Future Work} \label{conclusion}

In this work, we applied the cross-view contrastive learning to model the cross-view cooperative association in bundle recommendation. 
We introduced the cross-view contrastive learning to regularize the cross-view representations and proposed a simple, efficient, yet effective method CrossCBR, which significantly enhanced the SOTA performance of bundle recommendation on three public datasets. Various ablation and model studies demystified the working mechanism behind such huge performance leap.

Even though CrossCBR has achieved great performance, the study of contrastive learning on bundle or even general recommendation is still in its infancy, and several directions are promising in the future. First, model-based data augmentations, which can introduce both hard negative and diverse positive samples, should be helpful for further performance improvements. Second, more potential approaches are to be explored for modeling the cross-view cooperative association. Third, the cross-view contrastive learning paradigm is easy to be generalized to other similar tasks, as long as two distinctive views exit.


\section*{acknowledgement}
This research/project is supported by the Sea-NExT Joint Lab, and CCCD Key Lab of Ministry of Culture and Tourism.

\bibliographystyle{ACM-Reference-Format}
\bibliography{0_main}


\begin{thebibliography}{40}


\ifx \showCODEN    \undefined \def \showCODEN     #1{\unskip}     \fi
\ifx \showDOI      \undefined \def \showDOI       #1{#1}\fi
\ifx \showISBNx    \undefined \def \showISBNx     #1{\unskip}     \fi
\ifx \showISBNxiii \undefined \def \showISBNxiii  #1{\unskip}     \fi
\ifx \showISSN     \undefined \def \showISSN      #1{\unskip}     \fi
\ifx \showLCCN     \undefined \def \showLCCN      #1{\unskip}     \fi
\ifx \shownote     \undefined \def \shownote      #1{#1}          \fi
\ifx \showarticletitle \undefined \def \showarticletitle #1{#1}   \fi
\ifx \showURL      \undefined \def \showURL       {\relax}        \fi
\providecommand\bibfield[2]{#2}
\providecommand\bibinfo[2]{#2}
\providecommand\natexlab[1]{#1}
\providecommand\showeprint[2][]{arXiv:#2}

\bibitem[\protect\citeauthoryear{Bai, Zhou, Song, Qu, An, Li, and Gao}{Bai
  et~al\mbox{.}}{2019}]%
        {bai2019personalized}
\bibfield{author}{\bibinfo{person}{Jinze Bai}, \bibinfo{person}{Chang Zhou},
  \bibinfo{person}{Junshuai Song}, \bibinfo{person}{Xiaoru Qu},
  \bibinfo{person}{Weiting An}, \bibinfo{person}{Zhao Li}, {and}
  \bibinfo{person}{Jun Gao}.} \bibinfo{year}{2019}\natexlab{}.
\newblock \showarticletitle{Personalized Bundle List Recommendation}. In
  \bibinfo{booktitle}{\emph{{WWW}}}. \bibinfo{publisher}{{ACM}},
  \bibinfo{pages}{60--71}.
\newblock


\bibitem[\protect\citeauthoryear{Cao, Nie, He, Wei, Zhu, and Chua}{Cao
  et~al\mbox{.}}{2017}]%
        {cao2017embedding}
\bibfield{author}{\bibinfo{person}{Da Cao}, \bibinfo{person}{Liqiang Nie},
  \bibinfo{person}{Xiangnan He}, \bibinfo{person}{Xiaochi Wei},
  \bibinfo{person}{Shunzhi Zhu}, {and} \bibinfo{person}{Tat{-}Seng Chua}.}
  \bibinfo{year}{2017}\natexlab{}.
\newblock \showarticletitle{Embedding Factorization Models for Jointly
  Recommending Items and User Generated Lists}. In
  \bibinfo{booktitle}{\emph{{SIGIR}}}. \bibinfo{publisher}{{ACM}},
  \bibinfo{pages}{585--594}.
\newblock


\bibitem[\protect\citeauthoryear{Chang, Gao, He, Jin, and Li}{Chang
  et~al\mbox{.}}{2020}]%
        {chang2020bundle}
\bibfield{author}{\bibinfo{person}{Jianxin Chang}, \bibinfo{person}{Chen Gao},
  \bibinfo{person}{Xiangnan He}, \bibinfo{person}{Depeng Jin}, {and}
  \bibinfo{person}{Yong Li}.} \bibinfo{year}{2020}\natexlab{}.
\newblock \showarticletitle{Bundle Recommendation with Graph Convolutional
  Networks}. In \bibinfo{booktitle}{\emph{{SIGIR}}}.
  \bibinfo{publisher}{{ACM}}, \bibinfo{pages}{1673--1676}.
\newblock


\bibitem[\protect\citeauthoryear{Chang, Gao, He, Jin, and Li}{Chang
  et~al\mbox{.}}{2021}]%
        {chang2021bundle}
\bibfield{author}{\bibinfo{person}{Jianxin Chang}, \bibinfo{person}{Chen Gao},
  \bibinfo{person}{Xiangnan He}, \bibinfo{person}{Depeng Jin}, {and}
  \bibinfo{person}{Yong Li}.} \bibinfo{year}{2021}\natexlab{}.
\newblock \showarticletitle{Bundle Recommendation and Generation with Graph
  Neural Networks}.
\newblock \bibinfo{journal}{\emph{IEEE Transactions on Knowledge and Data
  Engineering}} (\bibinfo{year}{2021}).
\newblock


\bibitem[\protect\citeauthoryear{Chen, Liu, He, Gao, and Zheng}{Chen
  et~al\mbox{.}}{2019c}]%
        {chen2019matching}
\bibfield{author}{\bibinfo{person}{Liang Chen}, \bibinfo{person}{Yang Liu},
  \bibinfo{person}{Xiangnan He}, \bibinfo{person}{Lianli Gao}, {and}
  \bibinfo{person}{Zibin Zheng}.} \bibinfo{year}{2019}\natexlab{c}.
\newblock \showarticletitle{Matching User with Item Set: Collaborative Bundle
  Recommendation with Deep Attention Network.}. In
  \bibinfo{booktitle}{\emph{IJCAI}}. \bibinfo{pages}{2095--2101}.
\newblock


\bibitem[\protect\citeauthoryear{Chen, Kornblith, Norouzi, and Hinton}{Chen
  et~al\mbox{.}}{2020}]%
        {chen2020simple}
\bibfield{author}{\bibinfo{person}{Ting Chen}, \bibinfo{person}{Simon
  Kornblith}, \bibinfo{person}{Mohammad Norouzi}, {and}
  \bibinfo{person}{Geoffrey~E. Hinton}.} \bibinfo{year}{2020}\natexlab{}.
\newblock \showarticletitle{A Simple Framework for Contrastive Learning of
  Visual Representations}. In \bibinfo{booktitle}{\emph{{ICML}}}
  \emph{(\bibinfo{series}{Proceedings of Machine Learning Research},
  Vol.~\bibinfo{volume}{119})}. \bibinfo{publisher}{{PMLR}},
  \bibinfo{pages}{1597--1607}.
\newblock


\bibitem[\protect\citeauthoryear{Chen, Gu, Ren, He, Xie, Guo, Yin, and
  Zhang}{Chen et~al\mbox{.}}{2019a}]%
        {chen2019semi}
\bibfield{author}{\bibinfo{person}{Weijian Chen}, \bibinfo{person}{Yulong Gu},
  \bibinfo{person}{Zhaochun Ren}, \bibinfo{person}{Xiangnan He},
  \bibinfo{person}{Hongtao Xie}, \bibinfo{person}{Tong Guo},
  \bibinfo{person}{Dawei Yin}, {and} \bibinfo{person}{Yongdong Zhang}.}
  \bibinfo{year}{2019}\natexlab{a}.
\newblock \showarticletitle{Semi-supervised User Profiling with Heterogeneous
  Graph Attention Networks.}. In \bibinfo{booktitle}{\emph{IJCAI}},
  Vol.~\bibinfo{volume}{19}. \bibinfo{pages}{2116--2122}.
\newblock


\bibitem[\protect\citeauthoryear{Chen, Huang, Xu, Guo, Guo, Sun, Li, Pfadler,
  Zhao, and Zhao}{Chen et~al\mbox{.}}{2019b}]%
        {chen2019pog}
\bibfield{author}{\bibinfo{person}{Wen Chen}, \bibinfo{person}{Pipei Huang},
  \bibinfo{person}{Jiaming Xu}, \bibinfo{person}{Xin Guo},
  \bibinfo{person}{Cheng Guo}, \bibinfo{person}{Fei Sun}, \bibinfo{person}{Chao
  Li}, \bibinfo{person}{Andreas Pfadler}, \bibinfo{person}{Huan Zhao}, {and}
  \bibinfo{person}{Binqiang Zhao}.} \bibinfo{year}{2019}\natexlab{b}.
\newblock \showarticletitle{{POG:} Personalized Outfit Generation for Fashion
  Recommendation at Alibaba iFashion}. In \bibinfo{booktitle}{\emph{{KDD}}}.
  \bibinfo{publisher}{{ACM}}, \bibinfo{pages}{2662--2670}.
\newblock


\bibitem[\protect\citeauthoryear{Deng, Wang, Zhao, Zou, Wu, Tao, Fan, and
  Chen}{Deng et~al\mbox{.}}{2020}]%
        {deng2020personalized}
\bibfield{author}{\bibinfo{person}{Qilin Deng}, \bibinfo{person}{Kai Wang},
  \bibinfo{person}{Minghao Zhao}, \bibinfo{person}{Zhene Zou},
  \bibinfo{person}{Runze Wu}, \bibinfo{person}{Jianrong Tao},
  \bibinfo{person}{Changjie Fan}, {and} \bibinfo{person}{Liang Chen}.}
  \bibinfo{year}{2020}\natexlab{}.
\newblock \showarticletitle{Personalized Bundle Recommendation in Online
  Games}. In \bibinfo{booktitle}{\emph{{CIKM}}}. \bibinfo{publisher}{{ACM}},
  \bibinfo{pages}{2381--2388}.
\newblock


\bibitem[\protect\citeauthoryear{Ding, Ma, Wong, and Chua}{Ding
  et~al\mbox{.}}{2021}]%
        {ding2021DGSR}
\bibfield{author}{\bibinfo{person}{Yujuan Ding}, \bibinfo{person}{Yunshan Ma},
  \bibinfo{person}{Wai~Keung Wong}, {and} \bibinfo{person}{Tat{-}Seng Chua}.}
  \bibinfo{year}{2021}\natexlab{}.
\newblock \showarticletitle{Leveraging Two Types of Global Graph for Sequential
  Fashion Recommendation}. In \bibinfo{booktitle}{\emph{{ICMR}}}.
  \bibinfo{publisher}{{ACM}}, \bibinfo{pages}{73--81}.
\newblock


\bibitem[\protect\citeauthoryear{Gao, Yao, and Chen}{Gao et~al\mbox{.}}{2021}]%
        {gao2021simcse}
\bibfield{author}{\bibinfo{person}{Tianyu Gao}, \bibinfo{person}{Xingcheng
  Yao}, {and} \bibinfo{person}{Danqi Chen}.} \bibinfo{year}{2021}\natexlab{}.
\newblock \showarticletitle{SimCSE: Simple Contrastive Learning of Sentence
  Embeddings}. In \bibinfo{booktitle}{\emph{{EMNLP} {(1)}}}.
  \bibinfo{publisher}{Association for Computational Linguistics},
  \bibinfo{pages}{6894--6910}.
\newblock


\bibitem[\protect\citeauthoryear{Glorot and Bengio}{Glorot and Bengio}{2010}]%
        {glorot2010understanding}
\bibfield{author}{\bibinfo{person}{Xavier Glorot} {and} \bibinfo{person}{Yoshua
  Bengio}.} \bibinfo{year}{2010}\natexlab{}.
\newblock \showarticletitle{Understanding the difficulty of training deep
  feedforward neural networks}. In \bibinfo{booktitle}{\emph{{AISTATS}}}
  \emph{(\bibinfo{series}{{JMLR} Proceedings}, Vol.~\bibinfo{volume}{9})}.
  \bibinfo{publisher}{JMLR.org}, \bibinfo{pages}{249--256}.
\newblock


\bibitem[\protect\citeauthoryear{Gutmann and Hyv{\"{a}}rinen}{Gutmann and
  Hyv{\"{a}}rinen}{2010}]%
        {gutmann2010noise}
\bibfield{author}{\bibinfo{person}{Michael Gutmann} {and} \bibinfo{person}{Aapo
  Hyv{\"{a}}rinen}.} \bibinfo{year}{2010}\natexlab{}.
\newblock \showarticletitle{Noise-contrastive estimation: {A} new estimation
  principle for unnormalized statistical models}. In
  \bibinfo{booktitle}{\emph{{AISTATS}}} \emph{(\bibinfo{series}{{JMLR}
  Proceedings}, Vol.~\bibinfo{volume}{9})}. \bibinfo{publisher}{JMLR.org},
  \bibinfo{pages}{297--304}.
\newblock


\bibitem[\protect\citeauthoryear{He, Deng, Wang, Li, Zhang, and Wang}{He
  et~al\mbox{.}}{2020}]%
        {he2020lightgcn}
\bibfield{author}{\bibinfo{person}{Xiangnan He}, \bibinfo{person}{Kuan Deng},
  \bibinfo{person}{Xiang Wang}, \bibinfo{person}{Yan Li},
  \bibinfo{person}{Yong{-}Dong Zhang}, {and} \bibinfo{person}{Meng Wang}.}
  \bibinfo{year}{2020}\natexlab{}.
\newblock \showarticletitle{LightGCN: Simplifying and Powering Graph
  Convolution Network for Recommendation}. In
  \bibinfo{booktitle}{\emph{{SIGIR}}}. \bibinfo{publisher}{{ACM}},
  \bibinfo{pages}{639--648}.
\newblock


\bibitem[\protect\citeauthoryear{Hjelm, Fedorov, Lavoie{-}Marchildon, Grewal,
  Bachman, Trischler, and Bengio}{Hjelm et~al\mbox{.}}{2019}]%
        {hjelm2018learning}
\bibfield{author}{\bibinfo{person}{R.~Devon Hjelm}, \bibinfo{person}{Alex
  Fedorov}, \bibinfo{person}{Samuel Lavoie{-}Marchildon},
  \bibinfo{person}{Karan Grewal}, \bibinfo{person}{Philip Bachman},
  \bibinfo{person}{Adam Trischler}, {and} \bibinfo{person}{Yoshua Bengio}.}
  \bibinfo{year}{2019}\natexlab{}.
\newblock \showarticletitle{Learning deep representations by mutual information
  estimation and maximization}. In \bibinfo{booktitle}{\emph{{ICLR}}}.
  \bibinfo{publisher}{OpenReview.net}.
\newblock


\bibitem[\protect\citeauthoryear{Hu, He, Gao, and Zhang}{Hu
  et~al\mbox{.}}{2020}]%
        {hu2020modeling}
\bibfield{author}{\bibinfo{person}{Haoji Hu}, \bibinfo{person}{Xiangnan He},
  \bibinfo{person}{Jinyang Gao}, {and} \bibinfo{person}{Zhi{-}Li Zhang}.}
  \bibinfo{year}{2020}\natexlab{}.
\newblock \showarticletitle{Modeling Personalized Item Frequency Information
  for Next-basket Recommendation}. In \bibinfo{booktitle}{\emph{{SIGIR}}}.
  \bibinfo{publisher}{{ACM}}, \bibinfo{pages}{1071--1080}.
\newblock


\bibitem[\protect\citeauthoryear{Kingma and Ba}{Kingma and Ba}{2014}]%
        {kingma2014adam}
\bibfield{author}{\bibinfo{person}{Diederik~P Kingma} {and}
  \bibinfo{person}{Jimmy Ba}.} \bibinfo{year}{2014}\natexlab{}.
\newblock \showarticletitle{Adam: A method for stochastic optimization}.
\newblock \bibinfo{journal}{\emph{arXiv preprint arXiv:1412.6980}}
  (\bibinfo{year}{2014}).
\newblock


\bibitem[\protect\citeauthoryear{Li, Lu, Wang, Shi, Xie, Yang, Yang, Zhang, and
  Lin}{Li et~al\mbox{.}}{2021}]%
        {li2021package}
\bibfield{author}{\bibinfo{person}{Chen Li}, \bibinfo{person}{Yuanfu Lu},
  \bibinfo{person}{Wei Wang}, \bibinfo{person}{Chuan Shi},
  \bibinfo{person}{Ruobing Xie}, \bibinfo{person}{Haili Yang},
  \bibinfo{person}{Cheng Yang}, \bibinfo{person}{Xu Zhang}, {and}
  \bibinfo{person}{Leyu Lin}.} \bibinfo{year}{2021}\natexlab{}.
\newblock \showarticletitle{Package Recommendation with Intra- and
  Inter-Package Attention Networks}. In \bibinfo{booktitle}{\emph{{SIGIR}}}.
  \bibinfo{publisher}{{ACM}}, \bibinfo{pages}{595--604}.
\newblock


\bibitem[\protect\citeauthoryear{Li, Wang, Zhang, He, and Chua}{Li
  et~al\mbox{.}}{2022}]%
        {RGCL}
\bibfield{author}{\bibinfo{person}{Sihang Li}, \bibinfo{person}{Xiang Wang},
  \bibinfo{person}{An Zhang}, \bibinfo{person}{Xiangnan He}, {and}
  \bibinfo{person}{Tat-Seng Chua}.} \bibinfo{year}{2022}\natexlab{}.
\newblock \showarticletitle{Let Invariant Rationale Discovery Inspire Graph
  Contrastive Learning}. In \bibinfo{booktitle}{\emph{{ICML}}}.
\newblock


\bibitem[\protect\citeauthoryear{Li, Wang, He, Chen, Xiao, and Chua}{Li
  et~al\mbox{.}}{2020}]%
        {li2020hierarchical}
\bibfield{author}{\bibinfo{person}{Xingchen Li}, \bibinfo{person}{Xiang Wang},
  \bibinfo{person}{Xiangnan He}, \bibinfo{person}{Long Chen},
  \bibinfo{person}{Jun Xiao}, {and} \bibinfo{person}{Tat{-}Seng Chua}.}
  \bibinfo{year}{2020}\natexlab{}.
\newblock \showarticletitle{Hierarchical Fashion Graph Network for Personalized
  Outfit Recommendation}. In \bibinfo{booktitle}{\emph{{SIGIR}}}.
  \bibinfo{publisher}{{ACM}}, \bibinfo{pages}{159--168}.
\newblock


\bibitem[\protect\citeauthoryear{Liu, Chen, Li, Yu, McAuley, and Xiong}{Liu
  et~al\mbox{.}}{2021}]%
        {liu2021contrastive}
\bibfield{author}{\bibinfo{person}{Zhiwei Liu}, \bibinfo{person}{Yongjun Chen},
  \bibinfo{person}{Jia Li}, \bibinfo{person}{Philip~S Yu},
  \bibinfo{person}{Julian McAuley}, {and} \bibinfo{person}{Caiming Xiong}.}
  \bibinfo{year}{2021}\natexlab{}.
\newblock \showarticletitle{Contrastive self-supervised sequential
  recommendation with robust augmentation}.
\newblock \bibinfo{journal}{\emph{arXiv preprint arXiv:2108.06479}}
  (\bibinfo{year}{2021}).
\newblock


\bibitem[\protect\citeauthoryear{Logeswaran and Lee}{Logeswaran and
  Lee}{2018}]%
        {logeswaran2018efficient}
\bibfield{author}{\bibinfo{person}{Lajanugen Logeswaran} {and}
  \bibinfo{person}{Honglak Lee}.} \bibinfo{year}{2018}\natexlab{}.
\newblock \showarticletitle{An efficient framework for learning sentence
  representations}. In \bibinfo{booktitle}{\emph{{ICLR} (Poster)}}.
  \bibinfo{publisher}{OpenReview.net}.
\newblock


\bibitem[\protect\citeauthoryear{Oord, Li, and Vinyals}{Oord
  et~al\mbox{.}}{2018}]%
        {oord2018representation}
\bibfield{author}{\bibinfo{person}{Aaron van~den Oord}, \bibinfo{person}{Yazhe
  Li}, {and} \bibinfo{person}{Oriol Vinyals}.} \bibinfo{year}{2018}\natexlab{}.
\newblock \showarticletitle{Representation learning with contrastive predictive
  coding}.
\newblock \bibinfo{journal}{\emph{arXiv preprint arXiv:1807.03748}}
  (\bibinfo{year}{2018}).
\newblock


\bibitem[\protect\citeauthoryear{Qin, Wang, and Li}{Qin et~al\mbox{.}}{2021}]%
        {qin2021world}
\bibfield{author}{\bibinfo{person}{Yuqi Qin}, \bibinfo{person}{Pengfei Wang},
  {and} \bibinfo{person}{Chenliang Li}.} \bibinfo{year}{2021}\natexlab{}.
\newblock \showarticletitle{The World is Binary: Contrastive Learning for
  Denoising Next Basket Recommendation}. In
  \bibinfo{booktitle}{\emph{{SIGIR}}}. \bibinfo{publisher}{{ACM}},
  \bibinfo{pages}{859--868}.
\newblock


\bibitem[\protect\citeauthoryear{Rendle, Freudenthaler, Gantner, and
  Schmidt-Thieme}{Rendle et~al\mbox{.}}{2012}]%
        {rendle2012bpr}
\bibfield{author}{\bibinfo{person}{Steffen Rendle}, \bibinfo{person}{Christoph
  Freudenthaler}, \bibinfo{person}{Zeno Gantner}, {and} \bibinfo{person}{Lars
  Schmidt-Thieme}.} \bibinfo{year}{2012}\natexlab{}.
\newblock \showarticletitle{BPR: Bayesian personalized ranking from implicit
  feedback}.
\newblock \bibinfo{journal}{\emph{arXiv preprint arXiv:1205.2618}}
  (\bibinfo{year}{2012}).
\newblock


\bibitem[\protect\citeauthoryear{Rendle, Freudenthaler, and
  Schmidt{-}Thieme}{Rendle et~al\mbox{.}}{2010}]%
        {rendle2010factorizing}
\bibfield{author}{\bibinfo{person}{Steffen Rendle}, \bibinfo{person}{Christoph
  Freudenthaler}, {and} \bibinfo{person}{Lars Schmidt{-}Thieme}.}
  \bibinfo{year}{2010}\natexlab{}.
\newblock \showarticletitle{Factorizing personalized Markov chains for
  next-basket recommendation}. In \bibinfo{booktitle}{\emph{{WWW}}}.
  \bibinfo{publisher}{{ACM}}, \bibinfo{pages}{811--820}.
\newblock


\bibitem[\protect\citeauthoryear{Wang and Isola}{Wang and Isola}{2020}]%
        {wang2020understanding}
\bibfield{author}{\bibinfo{person}{Tongzhou Wang} {and}
  \bibinfo{person}{Phillip Isola}.} \bibinfo{year}{2020}\natexlab{}.
\newblock \showarticletitle{Understanding Contrastive Representation Learning
  through Alignment and Uniformity on the Hypersphere}. In
  \bibinfo{booktitle}{\emph{{ICML}}} \emph{(\bibinfo{series}{Proceedings of
  Machine Learning Research}, Vol.~\bibinfo{volume}{119})}.
  \bibinfo{publisher}{{PMLR}}, \bibinfo{pages}{9929--9939}.
\newblock


\bibitem[\protect\citeauthoryear{Wang, He, Wang, Feng, and Chua}{Wang
  et~al\mbox{.}}{2019}]%
        {wang2019neural}
\bibfield{author}{\bibinfo{person}{Xiang Wang}, \bibinfo{person}{Xiangnan He},
  \bibinfo{person}{Meng Wang}, \bibinfo{person}{Fuli Feng}, {and}
  \bibinfo{person}{Tat{-}Seng Chua}.} \bibinfo{year}{2019}\natexlab{}.
\newblock \showarticletitle{Neural Graph Collaborative Filtering}. In
  \bibinfo{booktitle}{\emph{{SIGIR}}}. \bibinfo{publisher}{{ACM}},
  \bibinfo{pages}{165--174}.
\newblock


\bibitem[\protect\citeauthoryear{Wei, Wang, Li, Nie, Li, Li, and Chua}{Wei
  et~al\mbox{.}}{2021}]%
        {wei2021contrastive}
\bibfield{author}{\bibinfo{person}{Yinwei Wei}, \bibinfo{person}{Xiang Wang},
  \bibinfo{person}{Qi Li}, \bibinfo{person}{Liqiang Nie}, \bibinfo{person}{Yan
  Li}, \bibinfo{person}{Xuanping Li}, {and} \bibinfo{person}{Tat{-}Seng Chua}.}
  \bibinfo{year}{2021}\natexlab{}.
\newblock \showarticletitle{Contrastive Learning for Cold-Start
  Recommendation}. In \bibinfo{booktitle}{\emph{{ACM} Multimedia}}.
  \bibinfo{publisher}{{ACM}}, \bibinfo{pages}{5382--5390}.
\newblock


\bibitem[\protect\citeauthoryear{Wu, Wang, Feng, He, Chen, Lian, and Xie}{Wu
  et~al\mbox{.}}{2021}]%
        {wu2021self}
\bibfield{author}{\bibinfo{person}{Jiancan Wu}, \bibinfo{person}{Xiang Wang},
  \bibinfo{person}{Fuli Feng}, \bibinfo{person}{Xiangnan He},
  \bibinfo{person}{Liang Chen}, \bibinfo{person}{Jianxun Lian}, {and}
  \bibinfo{person}{Xing Xie}.} \bibinfo{year}{2021}\natexlab{}.
\newblock \showarticletitle{Self-supervised Graph Learning for Recommendation}.
  In \bibinfo{booktitle}{\emph{{SIGIR}}}. \bibinfo{publisher}{{ACM}},
  \bibinfo{pages}{726--735}.
\newblock


\bibitem[\protect\citeauthoryear{Wu, Wang, Zhang, He, and Chua}{Wu
  et~al\mbox{.}}{2022}]%
        {DIR}
\bibfield{author}{\bibinfo{person}{Yingxin Wu}, \bibinfo{person}{Xiang Wang},
  \bibinfo{person}{An Zhang}, \bibinfo{person}{Xiangnan He}, {and}
  \bibinfo{person}{Tat-Seng Chua}.} \bibinfo{year}{2022}\natexlab{}.
\newblock \showarticletitle{Discovering Invariant Rationales for Graph Neural
  Networks}. In \bibinfo{booktitle}{\emph{ICLR}}.
\newblock


\bibitem[\protect\citeauthoryear{Xia, Yin, Yu, Shao, and Cui}{Xia
  et~al\mbox{.}}{2021a}]%
        {xia2021self}
\bibfield{author}{\bibinfo{person}{Xin Xia}, \bibinfo{person}{Hongzhi Yin},
  \bibinfo{person}{Junliang Yu}, \bibinfo{person}{Yingxia Shao}, {and}
  \bibinfo{person}{Lizhen Cui}.} \bibinfo{year}{2021}\natexlab{a}.
\newblock \showarticletitle{Self-Supervised Graph Co-Training for Session-based
  Recommendation}. In \bibinfo{booktitle}{\emph{{CIKM}}}.
  \bibinfo{publisher}{{ACM}}, \bibinfo{pages}{2180--2190}.
\newblock


\bibitem[\protect\citeauthoryear{Xia, Yin, Yu, Wang, Cui, and Zhang}{Xia
  et~al\mbox{.}}{2021b}]%
        {xia2020self}
\bibfield{author}{\bibinfo{person}{Xin Xia}, \bibinfo{person}{Hongzhi Yin},
  \bibinfo{person}{Junliang Yu}, \bibinfo{person}{Qinyong Wang},
  \bibinfo{person}{Lizhen Cui}, {and} \bibinfo{person}{Xiangliang Zhang}.}
  \bibinfo{year}{2021}\natexlab{b}.
\newblock \showarticletitle{Self-Supervised Hypergraph Convolutional Networks
  for Session-based Recommendation}. In \bibinfo{booktitle}{\emph{{AAAI}}}.
  \bibinfo{publisher}{{AAAI} Press}, \bibinfo{pages}{4503--4511}.
\newblock


\bibitem[\protect\citeauthoryear{Xie, Sun, Liu, Wu, Gao, Ding, and Cui}{Xie
  et~al\mbox{.}}{2020}]%
        {xie2020contrastive}
\bibfield{author}{\bibinfo{person}{Xu Xie}, \bibinfo{person}{Fei Sun},
  \bibinfo{person}{Zhaoyang Liu}, \bibinfo{person}{Shiwen Wu},
  \bibinfo{person}{Jinyang Gao}, \bibinfo{person}{Bolin Ding}, {and}
  \bibinfo{person}{Bin Cui}.} \bibinfo{year}{2020}\natexlab{}.
\newblock \showarticletitle{Contrastive Learning for Sequential
  Recommendation}.
\newblock \bibinfo{journal}{\emph{arXiv preprint arXiv:2010.14395}}
  (\bibinfo{year}{2020}).
\newblock


\bibitem[\protect\citeauthoryear{Ying, He, Chen, Eksombatchai, Hamilton, and
  Leskovec}{Ying et~al\mbox{.}}{2018}]%
        {ying2018graph}
\bibfield{author}{\bibinfo{person}{Rex Ying}, \bibinfo{person}{Ruining He},
  \bibinfo{person}{Kaifeng Chen}, \bibinfo{person}{Pong Eksombatchai},
  \bibinfo{person}{William~L. Hamilton}, {and} \bibinfo{person}{Jure
  Leskovec}.} \bibinfo{year}{2018}\natexlab{}.
\newblock \showarticletitle{Graph Convolutional Neural Networks for Web-Scale
  Recommender Systems}. In \bibinfo{booktitle}{\emph{{KDD}}}.
  \bibinfo{publisher}{{ACM}}, \bibinfo{pages}{974--983}.
\newblock


\bibitem[\protect\citeauthoryear{Yu, Yin, Gao, Xia, Zhang, and Hung}{Yu
  et~al\mbox{.}}{2021}]%
        {yu2021socially}
\bibfield{author}{\bibinfo{person}{Junliang Yu}, \bibinfo{person}{Hongzhi Yin},
  \bibinfo{person}{Min Gao}, \bibinfo{person}{Xin Xia},
  \bibinfo{person}{Xiangliang Zhang}, {and} \bibinfo{person}{Nguyen Quoc~Viet
  Hung}.} \bibinfo{year}{2021}\natexlab{}.
\newblock \showarticletitle{Socially-Aware Self-Supervised Tri-Training for
  Recommendation}. In \bibinfo{booktitle}{\emph{{KDD}}}.
  \bibinfo{publisher}{{ACM}}, \bibinfo{pages}{2084--2092}.
\newblock


\bibitem[\protect\citeauthoryear{Zhang, Hooi, Hu, Liang, and Feng}{Zhang
  et~al\mbox{.}}{2021}]%
        {zhang2021unleashing}
\bibfield{author}{\bibinfo{person}{Yifan Zhang}, \bibinfo{person}{Bryan Hooi},
  \bibinfo{person}{Dapeng Hu}, \bibinfo{person}{Jian Liang}, {and}
  \bibinfo{person}{Jiashi Feng}.} \bibinfo{year}{2021}\natexlab{}.
\newblock \showarticletitle{Unleashing the Power of Contrastive Self-Supervised
  Visual Models via Contrast-Regularized Fine-Tuning}. In
  \bibinfo{booktitle}{\emph{NeurIPS}}. \bibinfo{pages}{29848--29860}.
\newblock


\bibitem[\protect\citeauthoryear{Zhou, Ma, Zhang, Zhou, and Yang}{Zhou
  et~al\mbox{.}}{2021a}]%
        {zhou2021contrastive}
\bibfield{author}{\bibinfo{person}{Chang Zhou}, \bibinfo{person}{Jianxin Ma},
  \bibinfo{person}{Jianwei Zhang}, \bibinfo{person}{Jingren Zhou}, {and}
  \bibinfo{person}{Hongxia Yang}.} \bibinfo{year}{2021}\natexlab{a}.
\newblock \showarticletitle{Contrastive Learning for Debiased Candidate
  Generation in Large-Scale Recommender Systems}. In
  \bibinfo{booktitle}{\emph{{KDD}}}. \bibinfo{publisher}{{ACM}},
  \bibinfo{pages}{3985--3995}.
\newblock


\bibitem[\protect\citeauthoryear{Zhou, Wang, Zhao, Zhu, Wang, Zhang, Wang, and
  Wen}{Zhou et~al\mbox{.}}{2020}]%
        {zhou2020s3}
\bibfield{author}{\bibinfo{person}{Kun Zhou}, \bibinfo{person}{Hui Wang},
  \bibinfo{person}{Wayne~Xin Zhao}, \bibinfo{person}{Yutao Zhu},
  \bibinfo{person}{Sirui Wang}, \bibinfo{person}{Fuzheng Zhang},
  \bibinfo{person}{Zhongyuan Wang}, {and} \bibinfo{person}{Ji{-}Rong Wen}.}
  \bibinfo{year}{2020}\natexlab{}.
\newblock \showarticletitle{S3-Rec: Self-Supervised Learning for Sequential
  Recommendation with Mutual Information Maximization}. In
  \bibinfo{booktitle}{\emph{{CIKM}}}. \bibinfo{publisher}{{ACM}},
  \bibinfo{pages}{1893--1902}.
\newblock


\bibitem[\protect\citeauthoryear{Zhou, Sun, Liu, Zhang, and Miao}{Zhou
  et~al\mbox{.}}{2021b}]%
        {zhou2021selfcf}
\bibfield{author}{\bibinfo{person}{Xin Zhou}, \bibinfo{person}{Aixin Sun},
  \bibinfo{person}{Yong Liu}, \bibinfo{person}{Jie Zhang}, {and}
  \bibinfo{person}{Chunyan Miao}.} \bibinfo{year}{2021}\natexlab{b}.
\newblock \showarticletitle{SelfCF: A Simple Framework for Self-supervised
  Collaborative Filtering}.
\newblock \bibinfo{journal}{\emph{arXiv preprint arXiv:2107.03019}}
  (\bibinfo{year}{2021}).
\newblock


\end{thebibliography}
\end{document}